\documentclass{article}

\usepackage{amsmath}
\usepackage{amssymb}
\usepackage{stmaryrd}
\usepackage{tabularx}
\usepackage[applemac]{inputenc}
\usepackage{supertabular}

\newcommand{\syntaxdef}{::=}

\newcommand{\procdef}{\displaystyle \mathop{=}^{\mbox{\scriptsize def}}}

\newcommand{\declqvarp}[1]{\theta[#1]:}

\newcommand{\declcvarp}[1]{\kappa[#1]:}
\newcommand{\declcvar}[1]{\kappa[#1]}
\newcommand{\point}{\ \bullet}

\newcommand{\tensor}{\otimes}
\newcommand{\prefix}{.}

\newcommand{\seq}{\ ;}
\newcommand{\para}{\parallel}

\newcommand{\rect}{\mbox{$[ \hspace{-1pt} ]$}}
\newcommand{\cond}[2]{\rect #1 \rightarrow #2}
\newcommand{\bigcond}[3]{\mathop{\rect}_{#1}\hspace{5pt} #2 \rightarrow #3}
\newcommand{\restrict}[1]{|#1\ } 
\newcommand{\restrictg}[1]{|\{#1\}\ } 
\newcommand{\envoi}[1]{\ !#1\  }
\newcommand{\recep}[1]{\ ?#1\  }

\newcommand{\term}{\mbox{\it end}}
\newcommand{\stopproc}{\mbox{\it nil}}

\newcommand{\transition}{\hspace{10pt}\longrightarrow\hspace{10pt}}
\newcommand{\transitionse}{\longrightarrow} 
\newcommand{\atransition}[1]{\hspace{10pt}\underrightarrow{\hspace{10pt}#1\hspace{10pt}}\hspace{10pt}}
\newcommand{\atransitionse}[1]{\underrightarrow{\hspace{10pt}#1\hspace{10pt}}}
\newcommand{\tautransition}{\hspace{10pt}\underrightarrow{\hspace{10pt}\tau\hspace{10pt}}\hspace{10pt}}
\newcommand{\tautransitionse}{\underrightarrow{\hspace{10pt}\tau\hspace{10pt}}}
\newcommand{\dtransition}{\hspace{10pt}\underrightarrow{\hspace{10pt}\delta\hspace{10pt}}\hspace{10pt}}
\newcommand{\dtransitionse}{\underrightarrow{\hspace{10pt}\delta\hspace{10pt}}}
\newcommand{\ptransition}[1]{\hspace{10pt}\longrightarrow_{#1}\hspace{10pt}}
\newcommand{\ptransitionse}[1]{\longrightarrow_{#1}}
\newcommand{\divtransition}{\hspace{10pt}\dashrightarrow\hspace{10pt}}
\newcommand{\divtransitionse}{\dashrightarrow}

\newcommand{\contextestable}{\downarrow}

\newcommand{\ket}[1]{|#1\rangle}
\newcommand{\bra}[1]{\langle#1|}
\newcommand{\contexte}[4]{<#1, #2 = #3, #4 >}
\newcommand{\scontexte}[4]{/<#1, #2 = #3, #4 >}
\newcommand{\spcontexte}[1]{/ #1}
\newcommand{\contexteqcq}{<s, q = \ket{\psi}, f >}

\newcommand{\symbcontexteprob}{\boxplus}
\newcommand{\bigsymbcontexteprob}{\mathop\boxplus}
\newcommand{\contexteprob}[5]{\bigsymbcontexteprob_{#1}\!\!<#2, #3 = #4, #5 >}
\newcommand{\contexteprobqcq}{\bigsymbcontexteprob_{p_i}\!\!<s_i, q_i = \ket{\psi_i}, f_i>}

\newcommand{\scontexteprob}[5]{/\bigsymbcontexteprob_{#1}\!\!<#2, #3 = #4, #5 >}

\newcommand{\spcontexteprob}[2]{/\bigsymbcontexteprob_{#1}#2}
\newcommand{\spcontexteprobbin}[3]{/ #2\symbcontexteprob_{#1}#3}

\newcommand{\ensemble}[1]{\{#1\}}

\newcommand{\varpile}[1]{\mbox{Var$(#1)$}} 
\newcommand{\pileconcat}{|} 
\newcommand{\pileajout}{.}

\newcommand{\nattype}{\mbox{Nat}}
\newcommand{\qubittype}{\mbox{Qubit}}

\newcommand{\dom}[1]{\mbox{dom($#1$)}}

\newcommand{\surcharge}[2]{#1 \mbox{$\vartriangleleft \hspace{-7pt} -$} #2}

\newcommand{\domrestrict}[2]{#2 \vartriangleleft #1}

\newcommand{\tailleseq}[1]{\mbox{size}(#1)}

\newcommand{\n}{I\!\! N}
\newcommand{\transfu}{\mathcal U}
\newcommand{\obs}{\mathcal O}

\newcommand{\module}[1]{|#1|}

\newcommand{\sautdeligne}{\vspace{10pt}}

\renewcommand{\domrestrict}[2]{#1|_{#2}}
\renewcommand{\pileconcat}{\pileajout}
 
\title{ Toward a Quantum Process Algebra }
\author{Philippe {\sc Jorrand}\footnote{Philippe.Jorrand@imag.fr} ,
Marie {\sc Lalire}\footnote{Marie.Lalire@imag.fr}\\
Leibniz Laboratory\\
46, avenue Félix Viallet
38000 Grenoble, France}

\begin{document}

\maketitle

\begin{abstract}
Quantum computations operate in the quantum world. For their results to be useful in any way, there is an intrinsic necessity of cooperation and communication controlled by the classical world. As a consequence, full formal descriptions of algorithms making use of quantum principles must take into account both quantum and classical computing components and assemble them so that they communicate and cooperate.
This paper aims at defining a high level language allowing the description of classical and quantum programming, and their cooperation.
Since process algebras provide a framework to model cooperating computations and have well defined semantics, they have been chosen as a basis for this language.
Starting with a classical process algebra, this paper explains how to transform it for including quantum computation.
The result is a quantum process algebra with its operational semantics, which can be used to fully describe quantum algorithms in their classical context.
\end{abstract}

\section{ Introduction }

Quantum algorithms are often described by means of quantum gate networks. See \cite{RieffelPolak00Intro}
for an introduction to quantum computing, and \cite{Gruska99Book} or \cite{NielsenChuang00Book} for a full account of quantum computing and quantum information.
This has several drawbacks, for instance, gate networks do not allow descriptions of loops nor conditional execution of parts of networks. 
So as to overcome these difficulties, a few quantum programming languages have been developed, such as:
QCL \cite{Omer00QCL}, an imperative language which aims at simulating quantum programs,
qGCL \cite{Zuliani01These} which allows the construction of proved correct quantum programs through a refinement method, and QPL \cite{Selinger03QPL}, a functional language with a denotational semantics.
There is also the work of Alessandra Di Pierro and Herbert Wiklicky \cite{DiPierro01QCP}
who adapted constraint programming to quantum computation, with the purpose of defining a semantical framework for quantum programming. More recently, André Van Tonder has developed a quantum lamb\-da calculus \cite{Tonder03LambdaCalculus}, based on a simplified linear
lambda calculus.
  
Cooperation between quantum and classical computations is inherent in quantum algorithmics.
For example, the quantum computation part is in general probabilistic:
it produces a result which is checked
by a classical part and, if this result is not correct, the quantum computation has to be repeated.
Teleportation of a qubit state from Alice to Bob \cite{BennettBrassard93Teleportation} is another good example of this cooperation. Indeed, Alice carries out a measurement, the classical result of which (two bits) is sent to Bob, and Bob uses this result to determine which quantum transformation he must apply.
Moreover, initial preparation of quantum states and measurement of quantum results are two essential forms of interactions between the classical and quantum kinds of computations which the language must be able to express.
Process algebras are a good candidate for such a language since they provide a framework for modeling cooperating computations.
In addition, they have well defined semantics and permit the transformation of programs as well as the formal study and analysis of their properties.

Process algebras have already been used in the context of quantum programming
in \cite{NagGay02Verif}, where the authors have modeled a quantum cryptographic protocol and verified its correctness with a classical process algebra. 
Starting with a classical process algebra in section \ref{sectionClasProcAlg}, this paper explains the essential features of quantum computation in section \ref{sectionQuantComput} and "quantumizes" the initial process algebra in section \ref{sectionQuantProcAlg}. Examples of short quantum programs are given in section \ref{sectionExamples}.

\section{ A Classical Process Algebra }
\label{sectionClasProcAlg}

The classical process algebra chosen here is quite similar to CCS \cite{Milner89CommConc} and Lotos \cite{Bolognesi87IsoLotos}. Its syntax and semantics are given in 
appendix \ref{annexCPA}.
In this process algebra, communication among processes is the only basic action.
There is a distinction between value emission denoted
 $g \envoi v$, where $g$ is a communication  gate and $v$ a value, and value reception denoted $g \recep x$, where $g$ is a gate and $x$ a variable which receives the value.
To create a process from basic actions, the prefix operator "$\prefix$" is used: if $\alpha$ is an action and $P$, a process, $\alpha\prefix P$ is a new process which performs $\alpha$ first, then behaves as $P$.

There are two predefined processes. The first one is $\stopproc$, the process that cannot perform any transition, and the other one is $\term$, which performs a "$\delta$-transition" for signaling successful termination, and becomes $\stopproc$ ("$\delta$-transitions" are necessary in the semantics of sequential composition of processes).

The operators of the process algebra are: sequential composition ($P\seq Q$), parallel composition ($P\para Q$), conditional choice ($\bigcond{i}{c_i}{P_i}$) and restriction ($P\restrict L$). As for sequential composition, process $Q$ is executed if process $P$ terminates successfully, that is to say if $P$ performs a $\delta$-transition.
The process $\bigcond{i}{c_i}{P_i}$, where $c_i$ is a condition and $P_i$ a process, evolves as a process chosen nondeterministically among the processes $P_j$ such that $c_j$ is true.
Restriction is useful for disallowing the use of some gates (the gates listed in $L$),
thus forcing internal communication within process $P$. 
Communication can occur between two parallel processes whenever a value emission in one of them and a value reception in the other one use the same gate name. For instance, a communication can occur in the process $g\envoi v\prefix P\para g\recep x \prefix Q$  on gate $g$. After the communication has occurred, this process  becomes $P\para Q[x\leftarrow v]$ where $Q[x\leftarrow v]$ is $Q$ where all occurrences of $x$ have been replaced by $v$.

\section{ Quantum Computing }
\label{sectionQuantComput}

\subsection{ Qubits and Registers }

Whereas classical computing is based on bits taking values in $\ensemble{0,1}$,
quantum computing is based on qubits (quantum bits). The states of qubits are normalized vectors in a two dimensional space $\mathbb C^2$, where $\mathbb C$ are the complex numbers.

In the following, we will use Dirac's notation $\ket .$ for vectors. 
Let $\ket 0$ and $\ket 1$ be two normalized vectors forming an orthonormal basis of
$\mathbb C^2$:
$$\ket 0 = \left(\begin{array}{c} 1\\ 0\end{array}\right)\hspace{20pt}
\ket 1 = \left(\begin{array}{c} 0\\ 1\end{array}\right)$$
The state of a qubit can be written $\ket\psi = \alpha \ket 0 + \beta \ket 1$ where
$\alpha, \beta\in\mathbb C $ and
$\module\alpha^2+\module\beta^2=1$ so that $\ket\psi$ is normalized.

Like in classical computing where bits are organized into registers, there are also registers of qubits. The state $\ket\psi$ of a register made of two qubits in
states $\ket{\psi_1}$ and $\ket{\psi_2}$ respectively, is the tensor product of the states of these qubits, that is to say $\ket\psi = \ket{\psi_1}\tensor\ket{\psi_2}$.
The state of a two qubit register is thus a normalized vector in a 4 dimensional space
$\mathbb C^4$ with basis $\ensemble{\ket i \otimes \ket j},\ i,j\in\ensemble{0,1}$,
usually denoted $\{\ket{00}, \ket{01}, \ket{10}, \ket{11}\}$:
$\ket\psi = \alpha\ket{00}+\beta\ket{01}+\gamma\ket{10}+\delta\ket{11}$.
This can be generalized to registers of $n$ qubits: their states are normalized vectors in a $2^n$ dimensional space $\mathbb C^{2^n}$.

It is important to note that the state of a $n$ qubit  register cannot in general be  written as a tensor product of the states of the qubits which compose the register.
Such states are "entangled" states.
For instance, the famous state
$\ket{EPR} = \frac{1}{\sqrt 2}(\ket{00}+\ket{11})$\footnotemark is an entangled state because for all $\alpha, \beta, \gamma, \delta$ in $\mathbb C$, $\frac{1}{\sqrt 2}(\ket{00}+\ket{11}) \neq (\alpha\ket 0 + \beta\ket 1)\tensor(\gamma\ket 0 + \delta\ket 1)$.

\footnotetext{EPR comes from Einstein, Podolsky and Rosen \cite{Einstein35EPR}, who questioned the completeness of quantum theory because of the existence of such states. Later, theoretical results \cite{Bell64EPR} followed by experiments \cite{Aspect82BellIneq} proved them wrong.}

It must be noted that entangled states are an essential difference between the classical and the quantum worlds: in the quantum world, it is not true in general that the state of a system composed of $n$ sub-systems can be reduced to an $n$-tuple of the states of its components. This will have to be taken into account in the semantics of the process algebra presented here.

\subsection{ Deterministic Evolution }

According to the postulates of quantum mechanics, the evolution of a closed quantum system — {\it i.e.} which is not observed — can be described by a unitary transformation.
A unitary transformation on $n$ qubits can be represented by a complex and unitary $2^n\times 2^n$ matrix $U$.
The unitarity condition maintains the normalization and can be written $UU^\dagger = U^\dagger U = I$ where $U^\dagger$ is the adjoint ({\it i.e.} conjugate transpose) of $U$.
This condition implies that the evolution of a closed quantum system is reversible.

\subsection{ Probabilistic Measurement }
\label{subsecProbMeas}

Measurement corresponds to the observation of a quantum system by a classical system.
Contrary to the classical world, where reading a bit does not change its state, the observation of qubits is destructive: in general, measuring a qubit modifies its state irreversibly.

Moreover, measurement is probabilistic: measuring two qu\-bits initially prepared in identical states, will not necessarily produce identical results. 
Let $\ket\psi = \alpha\ket 0 + \beta\ket 1$ be the state of a qubit.
Measurement of a qubit is performed relatively
to a basis of $\mathbb C^2$.
Measuring this qubit in the standard basis $\ensemble{\ket 0, \ket 1}$, yields $\ket 0$ with probability $\module{\alpha}^2$, and $\ket 1$ with probability $\module{\beta}^2$.
 For this reason, unlike unitary transformations, measurement is not reversible.

Only so-called "projective" measurements will be considered here.
A projective measurement is described by an hermitian matrix $M$ ({\it i.e.} $M = M^\dagger$), called an observable. 
The set of observables is denoted $\obs$.
Quantum registers of $n$ qubits measured with observable $M$ (a $2^n\times2^n$ matrix) have their state projected onto one of the eigenspaces of $M$ and renormalized. 
Indeed, since $M$ is hermitian, it has a spectral decomposition, $M= \sum_m mP_m$ where $P_m$ is the projector onto the eigenspace of $M$ corresponding to the eigenvalue $m$ (which is real since $M$ is hermitian). So, if a quantum register in state $\ket\psi$ is measured with observable $M$, its state is transformed to:
$$\frac{P_m\ket\psi}{\sqrt{p_m}}$$
with probability $p_m$.
$P_m\ket\psi$ is the projection of $\ket\psi$ onto the eigenspace corresponding to $m$, and 
$\sqrt{p_m}$ is the renormalization factor.  The probability $p_m$ is $\bra\psi P_m \ket\psi$,
the scalar product of $\ket\psi$ and $P_m\ket\psi$.
The result of the measurement, as viewed by the classical observer, is the value $m$.

For example, the observable $M_{std}$ for measuring one qubit in the standard basis, is :
$$M_{std} = 0 P_0+1 P_1 = 
\left( \begin{array}{cc} 0&0\\ 0&1\end{array}\right)$$
where $P_0$ and $P_1$ are the outer products 
$\ket 0\bra 0 = \left( \begin{array}{cc} 1&0\\ 0&0\end{array}\right)$ and
$\ket 1\bra 1 = \left( \begin{array}{cc} 0&0\\ 0&1\end{array}\right)$ respectively
(with Dirac's notation, $\bra 0 = ( \begin{array}{cc} 1&0\end{array})$,
$\bra 1 = ( \begin{array}{cc} 0&1\end{array})$).

\subsection{ Two Important Quantum Features }

\subsubsection{ Entangled States. }
\label{subsubsecEntangledStates}
Entangled states show amazing properties, especially when measurements are performed on parts of them.
For instance, with a two qubit register in state $\ket{EPR} = \frac{1}{\sqrt 2}(\ket{00}+\ket{11})$,
measuring one of the two qubits in the standard basis yields:
\begin{itemize}
\item either 0, with probability $1/2$, and the state becomes $\ket{00}$,
\item or 1, with probability $1/2$, and the state becomes $\ket{11}$.
\end{itemize}
After that, if the other qubit is also measured in the standard basis, the same result as for the first measurement is obtained with probability $1$: the results of the two measurements are correlated and this is independent of the spatial arrangement of the qubits and the distance separating them.

\subsubsection{ No Cloning Theorem. }
An other important difference between quan\-tum computing and classical computing is that in the quantum world, it is impossible to copy the state of quantum systems. This is known as the {\it no cloning theorem}: an unknown quantum state cannot be cloned. In the context of languages and programming, this means that it is impossible to copy the value of a quantum variable into another quantum variable.

\section{ "Quantumized" Processes }
\label{sectionQuantProcAlg}

\subsection{ Quantum Variables }
 \label{subsecQuantVar}

There are two types of variables in the "quantumized" process algebra, one classical: {\it $\nattype$}, for variables taking integer values, and one quantum: {\it $\qubittype$} for variables standing for qubits.

In classical process algebras, variables are instantiated when communications between processes  occur and cannot be modified after their instantiation. As a consequence, it is not necessary to store their values. In fact, when a variable is instantiated, all its occurrences are replaced by the value received (see the semantics of communication in parallel composition, as given in appendix \ref{annexCPA}).

Here, quantum variables stand for physical qubits. Applying a unitary transformation to a variable which represents a qubit modifies the state of that qubit.
This means that values of variables are modified. For that reason, it is necessary to keep track of both variable names and variable states.

Since variables are no longer just names standing for communicated values, they have to be declared. The syntax of declarations is:
$$\declcvar{x_1,\ldots,x_n}\declqvarp{y_1,\ldots,y_m} P\point$$
 where  $x_1,\ldots,x_n$ is a list of classical variables, $y_1,\ldots,y_n$ is a list of quantum variables, and $P$ is a process which can make use of these classical and quantum variables.
Variables have their scope limited on the left by their declaration and on the right by the postfixed operator $\point$.
To simplify the rest of this paper, the names of variables will always be considered distinct. 

As already said, it is necessary to store the states and the names of variables during the execution of a process. Consequently, in the inference rules which describe the semantics of processes, the states of processes can no longer be process terms only,
as it was the case for the classical process algebra,
they have to be process terms $P$ together with contexts $C$, of the form $P\spcontexte C$.
The main purpose of a context is to maintain the quantum state,
stored as $q = \ket\psi$ where $q$ is a sequence of quantum variable names
and $\ket\psi$ their quantum state. 
Moreover, in order to treat classical variables in a similar way, 
modifications of classical variables are also allowed. So, for the same reason as in the case of quantum variables, classical values are stored in the context.
Storing and retrieving classical values is represented by functions
$f: \mbox{\it names} \rightarrow\mbox{\it values}$.
The context must also keep track of the embedding of variable scopes.
To keep track of parallel composition,
this is done via a "cactus stack" structure of sets of variables, called the environment stack ($s$), which stores variable scopes and types. The set of all the variables in $s$ is denoted $\varpile s$.

In summary, the context has three components $\contexteqcq$, where:
\begin{itemize}
\item $s$ is the environment stack;
\item $q$ is a sequence of quantum variable names;
\item $\ket\psi$ is the quantum state of the variables in $q$;
\item $f$ is the function which associates values to classical variables.
\end{itemize}

\sautdeligne
The rules for declaration and liberation of variables are the following:

{\bf Declaration:}
$$
\frac{}{\declcvar{x_1,\ldots,x_n}\ \declqvarp{y_1,\ldots,y_m}  P \point \spcontexte{C}
\transition P \point \spcontexte{C'}}
$$
with $C = \contexteqcq$, $C' = \contexte{s'}{q}{\ket\psi}{f}$ and\\
$s' = \ensemble{(x_1,\nattype),\ldots,(x_n,\nattype),
(y_1,\qubittype),\ldots,(y_m,\qubittype)}\pileajout s$

\sautdeligne
\noindent
This rule adds the new variable names and types to the stack $s$. Because the variables do not have values yet, the quantum state and the classical function do not have to be modified at this point.

{\bf Evolution of a process within the scope of declared variables:}
$$
\frac{P \spcontexte{C} \divtransition P' \spcontexte{C'}}
{P \point \spcontexte{C} \divtransition P'\point \spcontexte{C'}}
$$
where $\divtransitionse$ stands for any of the transitions: $\atransitionse\alpha$ with $\alpha$ an action, $\tautransitionse$ with $\tau$ the "silent" action,  and the declaration transition $\transitionse$.

In short:
if the process $P$ can perform a transition, then the process $P\point$
can perform the same transition, provided that the action of the transition is not $\delta$. 

{\bf Termination of a process with exit from a scope and liberation of the variables:}
$$
\frac{P \spcontexte{C} \dtransition P' \scontexte{e \pileajout s}{q}{\ket{\psi}}{f}}
{P \point \spcontexte{C} \dtransition \stopproc\ 
\scontexte{s}{q[e\leftarrow *]}{\ket\psi}{\domrestrict{f}{\varpile s}}}
$$

If the action is $\delta$, this means that $P$ has successfully terminated, so the context must be cleaned up by eliminating the variables having their scope limited to that process. 

Cleaning up the context means eliminating the head of the stack and restricting the function $f$ to the variables remaining in the stack ($\domrestrict{f}{E}$ means $f$ restricted to $E$). As regards to the quantum part of the context, because of possible entanglement among local variables and other more global ones, qubits corresponding to these local variables cannot be removed. Only their variable names are erased and replaced by a "$*$" in the sequence $q$ ($q[e\leftarrow *]$ is $q$ in which all the names listed in $e$ have been replaced by $*$).
The quantum state is not modified.

\subsection{ Basic Actions }

The classical  basic actions are classical to classical communications. Classical to quantum communications are introduced for initializing qubits. Quantum to classical communications are part of measurement and are dealt with in the next paragraph.

The semantics of communications is based upon the following rules:
$$
\frac{}{g\envoi{v} \prefix P \spcontexte{C} \atransition{g\envoi{v}} P \spcontexte{C}} 
\hspace{10pt} v \in \n 
$$
$$
\frac{}{g\recep x \prefix P \spcontexte C \atransition{g\recep x} P \spcontexte C}
$$
with $C=\contexteqcq$, $x\in \varpile s$, and $x\not\in q$. 

The first rule deals with classical value sending, and the second one, with value reception.
It should be noted that in the second rule, the variable $x$ can be classical or quantum
but, if it is quantum, it must not have already been initialized.
In the semantics of parallel composition, the combination of these rules defines communication.
If $x$ is a qubit, the communication initializes it in the basis state $\ket v$, where $v$ is the classical value sent (in this case, $v$ must be $0$ or $1$).

\sautdeligne
The second kind of basic actions is unitary transformations which perform the unitary evolution of qubit states. Given a set $\transfu$ of predefined unitary transformations, the action corresponding to the application of $U\in\transfu$ to a list of quantum variables is denoted by $U[x_1,\ldots,x_n]$. 

The inference rule for unitary transformations is:
$$
\frac{}{U[x_1,\ldots,x_n]\prefix P \spcontexte C \tautransition
P \spcontexte{C'}}
$$
where
\begin{itemize}
\item $C=\contexteqcq$, $C'=\contexte{s}{q}{\ket{\psi'}}{f}$
\item  $U\in \transfu$, $x_1,\ldots,x_n \in \varpile s$, and $x_1,\ldots,x_n \in q$ 
\item $\ket{\psi'} = \Pi^t.(U\otimes I^{\otimes k}).\Pi\ket\psi$
\item $\Pi$ is the permutation matrix which places the $x_i$'s at the head of $q$
and $\Pi^t$ is the transpose of $\Pi$
\item $k = \tailleseq q - n\ $
\item$I^{\otimes k}\!= \underbrace{I \otimes \cdots \otimes I}_k$, where $I$ is the identity matrix on $\mathbb C^2$
\end{itemize}

The condition $x_1,\ldots,x_n \in q$ prevents from applying a unitary transformation to qubits which have not been initialized.
The third point deals with the evolution from a quantum state initially equal to $\ket\psi$.
Since the unitary transformation $U$ may be applied to qubits which are anywhere within the list $q$, a permutation $\Pi$ must be applied first. This permutation moves the $x_i$'s so that they are placed at the head of $q$ in the order specified by $[x_1,\ldots,x_n]$.
Then $U$ can be applied to the first $n$ elements and $I$ to the remainder. Finally, the last operation is the inverse of the permutation $\Pi$ ($\Pi^{-1} = \Pi^t$) so that at the end, the elements in $q$ and $\ket\psi$ are put back in the same order.

\subsection{ Measurement and Probabilistic Processes }

A last but essential basic action has to be introduced into the process algebra: quantum measurement.
Let $M\in\obs$ be an observable, $x_1,\ldots,x_n$ a list of quantum variables and $g$ a gate.
Then, the syntax for measurement is the following:
\begin{itemize}
\item $M[x_1,\ldots,x_n]$ is a measurement of the $n$ qubits of the list with respect to observable $M$, but the classical result is neither stored nor transmitted.
\item $g \envoi{M[x_1,\ldots,x_n]}$ is a measurement of the $n$ qubits of the list with respect to observable $M$, followed by sending the classical result through gate $g$.
\end{itemize}

As said in paragraph \ref{subsecProbMeas}, measurement is probabilistic: more precisely, the classical result and the quantum state after measurement are probabilistic.
This requires the introduction of a probabilistic composition operator for contexts.
This operator is denoted $\symbcontexteprob_p$:
the state $P\spcontexteprobbin{p}{C_1}{C_2}$ is $P\spcontexte{C_1}$ with probability $p$ and $P\spcontexte{C_2}$ with probability $1-p$.

This implies that, in general, the context is either of the form $\contexteqcq$, or of the form
$\contexteprobqcq$ where the $p_i$'s are probabilities.

As explained in \cite{Cazorla01Art,CazorlaMis}, if a process contains both a probabilistic and a nondeterministic choice, the probabilistic choice must always be solved first.
In the process algebra presented here, nondeterminism appears with parallel composition and conditional choice. So as to guarantee that probabilistic choice is always solved first, the notion of probabilistic stability for contexts is introduced: a context $C$ is probabilistically stable, which is denoted $C\contextestable$, if it is of the form $\contexteqcq$.
If the context of a process state is not stable, this state must perform a probabilistic transition.

The semantic rule for measurement without communication is:
$$
\frac{}{M[x_1,\ldots,x_n]\prefix P \spcontexte C \tautransition
P\scontexteprob{p_i}{s}{q}{\ket{\psi_i}}{f}}
$$
with
\begin{itemize}
\item $C=\contexteqcq$ (which implies $C\contextestable$)
\item $x_1,\ldots,x_n \in \varpile s$ and $x_1,\ldots,x_n \in q$ 
\item  $M\in\obs$  with $\sum_{i} \lambda_i P_i$ as spectral decomposition
\item $p_i = \langle \psi | \Pi^t (P_i \otimes I^{\otimes k}) \Pi | \psi \rangle$
\item $\displaystyle \ket{\psi_i} =  \frac{\Pi^t (P_i \otimes I^{\otimes k}) \Pi \ket{\psi}}{\sqrt{p_i}}$
\item $\Pi$ is the permutation matrix which places the $x_i$'s at the head of $q$
and $\Pi^t$ is the transpose of $\Pi$
\item $k = \tailleseq q - n\ $
\end{itemize}

As in the case of unitary transformations, a permutation $\Pi$ rearranges the qubits so that projectors apply only to measured qubits.
The computations of $\ket{\psi_i}$ and $p_i$ stem from the projective measurement postulate of quantum mechanics as summarized in paragraph \ref{subsecProbMeas}.

When the value coming out of the measurement is sent out, the rule is:
$$
\frac{}{g\envoi{M[x_1,\ldots,x_n]} \prefix P \spcontexte C \tautransition
(g\envoi y\prefix \term \point)\seq P\spcontexteprob{p_i}{C_i}}
$$
where
\begin{itemize}
\item $y$ is a new variable (implicitly declared as $\declcvar y$, see below)
\item $C = \contexteqcq$ (which implies $C\contextestable$)
\item $C_i = \contexte{\ensemble{(y,\nattype)}\pileajout s}{q}{\ket{\psi_i}}
{\surcharge f \ensemble{y\mapsto\lambda_i}}$
\item and the conditions are the same as in the rule without communication.
\end{itemize}

\sautdeligne
The only remaining point is the evolution of processes with probabilistic contexts. It is necessary to introduce probabilistic transitions for describing this evolution: $S_1 \ptransition p S_2$ means that state $S_1$ becomes $S_2$ with probability $p$. This is used in the following rule:
$$
\frac{}{P\spcontexteprob{p_i}{C_i} \ptransition{p_i} P\spcontexte{C_i}}
\mbox{ where } \sum_j p_j = 1
$$

The syntax and the main inference rules of this quantum process algebra are presented in appendix \ref{annexQPA}.

\section{ Examples }
\label{sectionExamples}

A few unitary transformations are often used in quantum algorithms:
\begin{itemize}
\item {\it Hadamard} is a transformation on one qubit denoted $H$.
Its action is the creation of uniform superpositions:

\noindent
$H : \ket 0 \mapsto \frac{1}{\sqrt 2}(\ket 0+\ket 1)$ and $\ket 1 \mapsto\frac{1}{\sqrt 2}(\ket 0-\ket 1)$ 

\item {\it Controlled Not} is a transformation on two qubits denoted $CNot$. If the first qubit is in state $\ket 1$, it flips the state of the second qubit:

\noindent
$CNot : \ket{00}\mapsto\ket{00},\ket{01}\mapsto\ket{01},
\ket{10}\mapsto\ket{11},\ket{11}\mapsto\ket{10}$

\item {\it Pauli matrices} are four transformations on one qubit:

\noindent
$$I = \left(\begin{array}{cc}1&0\\ 0&1\end{array}\right),
X = \left(\begin{array}{cc}0&1\\ 1&0\end{array}\right),$$
$$Y = \left(\begin{array}{cc}0&-i\\ i&0\end{array}\right),
Z = \left(\begin{array}{cc}1&0\\ 0&-1\end{array}\right)$$
\end{itemize}

\subsection{Construction of an EPR pair}

\newcommand{\eprinit}{\mbox{BuildEPR}}
\newcommand{\verifepr}{\mbox{CheckEPR}}
$$
\begin{array}{lcl}
\eprinit & \procdef & \declqvarp{x,y}\\
&&(( g_1\recep{x} \prefix g_2 \recep y \prefix H[x] \prefix CNot[x,y] \prefix\term)\\
&&\para (g_1\envoi 0\prefix g_2\envoi 0\prefix\term))\\
&&\restrictg{g_1,g_2} \point\\
\end{array}
$$
This process puts the pair of qubits $x, y$ in the state $\ket{EPR}$
(see paragraph \ref{subsubsecEntangledStates}).
To check that the order of measurement of the two qubits does not matter, it is possible, using the inference rules, to analyze the behaviour of the following two processes: in both of them, the first measurement produces $0$ ($1$) with probability $0.5$ and the second measurement produces
$0$ ($1$) with probability 1.

$$
\begin{array}{rcll}
\verifepr_1 &\procdef&\declqvarp{a,b}& \eprinit[a,b]\seq\\
&&& M_{std}[a] \prefix M_{std}[b]\prefix \term\point\\
&&&\\
\verifepr_2 &\procdef&\declqvarp{a,b}& \eprinit[a,b]\seq\\
&&& M_{std}[b] \prefix M_{std}[a]\prefix \term\point
\end{array}
$$

\subsection{Teleportation}

Once upon a time, there were two friends, Alice and Bob who had to separate and live away from each other.
Before leaving, each one took a qubit of the same EPR pair.
Then Bob went very far away, to a place that Alice did not know.
Later on, someone gave Alice a mysterious qubit in a state $\ket\psi = \alpha\ket 0+\beta\ket 1$,
with a mission to forward this state to Bob.
Alice could neither meet Bob and give him the qubit, nor clone it and broadcast copies everywhere, nor measure it to know $\alpha$ and $\beta$. Nevertheless, Alice succeeded thanks to the EPR pair and the teleportation protocol \cite{BennettBrassard93Teleportation}:

\newcommand{\alice}{\mbox{Alice}}
\newcommand{\bob}{\mbox{Bob}}
\newcommand{\teleport}{\mbox{Teleport}}
\newcommand{\meas}{\mbox{\it meas}}

$$
\begin{array}{lcl}
\alice&\procdef& \declqvarp{x,y} CNot[x,y] \prefix H[x] \prefix
\meas \envoi{M[x,y]} \prefix \term \point\\
&&\\
\bob&\procdef&\declqvarp{z}(\declcvarp{k} \meas\recep k\prefix
                      \cond{k=0}{I[z]\prefix\term}\\
                      && \hspace{94pt}\cond{k=1}{X[z]\prefix\term}\\
                      && \hspace{94pt}\cond{k=2}{Z[z]\prefix\term}\\
                      && \hspace{94pt}\cond{k=3}{Y[z]\prefix\term}\point)\point\\
&&\\
\teleport&\procdef& \declqvarp{\psi}  (\declqvarp{a,b} \eprinit[a,b]\seq\\
&& \hspace{31pt}  (\alice[\psi,a]\para\bob[b])\restrictg\meas\point)\point\\
\end{array}
$$

$M$ is the observable corresponding to measuring two qubits in the standard basis of
$\mathbb C^4$:
$$
M =
\left(\begin{array}{cccc}
0&0&0&0\\
0&1&0&0\\
0&0&2&0\\
0&0&0&3\\
\end{array}\right)
$$

The inference rules can be used to show that this protocol results in Bob's $z$ qubit having the state initially possessed by the $x$ qubit of Alice, with only two classical bits sent from Alice to Bob.

\section{ Conclusion }

This paper has presented a process algebra for quantum programming.  One of its advantages is that it can describe classical and quantum programming, and their cooperation. Without this cooperation, the implementation of the teleportation protocol, for instance, is not possible. Another feature of this language is that measurement and initialization of quantum registers appear through communications between quantum and classical parts of the language, which happens to be a faithful model of physical reality.

Moreover, a thorough semantics has been defined, thus allowing the study and analysis of programs. One peculiarity of  this  semantics is the introduction of probabilistic processes, due to quantum measurement. Probabilistic processes perform probabilistic transitions. As a consequence, the execution tree obtained  from a process presents action and probabilistic branches.

Several extensions are possible. Firstly, quantum to quantum communications could be added to allow the modeling of cryptographic protocols.
Another track that could be followed is the use of density matrices, which are a more general description of quantum states than vectors in $\mathbb C^{2^n}$. Density matrices notably permit to describe states of parts of entangled registers. They would give a more abstract semantics to this process algebra and open the way to a semantic analysis similar to abstract interpretation.

\section{ Acknowledgment }

The authors thank Frédéric Prost for having suggested nice improvements to a previous version of this work.

\bibliographystyle{abbrv}
\bibliography{biblio}

\appendix
\section{A classical process algebra}
\label{annexCPA}
\subsection{Syntax of process terms}

\begin{tabular}{lcl}
\it elem\_cond & $\syntaxdef$ &\it variable $|$ value \\
\it comp & $\syntaxdef$ & $\pmb= | \pmb\neq | \pmb\le | \pmb\ge | \pmb< | \pmb>$ \\
\it cond & $\syntaxdef$ & \it elem\_cond comp elem\_cond\\
&&\\
\it communication & $\syntaxdef$ & \it gate {\bf !} value $|$ gate {\bf ?} variable  \\
&&\\
\it process & $\syntaxdef$ & \it $\pmb\stopproc$\\
                  &$|$& \it $\pmb\term$ \\
                  &$|$& \it process\_name \\
                  &$|$& \it communication $\pmb\prefix$ process\\
                  &$|$& \it  process $\pmb\seq$ process\\
                  &$|$& \it  process $\pmb\para$ process\\
                  &$|$& \it  $\{\pmb\rect$ cond $\pmb\rightarrow$ process $\}^+$\\
                  &$|$& \it  process $\pmb | \pmb\{$ gate $\{$ ,gate $\}^*\pmb\}$ \\
&&\\
\it proc\_decl & $\syntaxdef$ & \it process\_name $\displaystyle\pmb{\procdef}$ process\\
\end{tabular}

\subsection{Semantics}
\label{annexSemCPA}

The semantics is specified with inference rules which give the evolution of the states of processes. In the classical process algebra considered here, the state of a process is a process term. The inference rules are of the form:
$$\frac{\mbox{\it Premises}}{\mbox{\it Conclusion}}\hspace{10pt} \mbox{\it Condition}$$
which means that if {\it Premises} have been established and {\it Condition} holds then the 
{\it Conclusion} can be infered.

{\it Premises} and {\it Conclusion} are of the form $P \divtransition P'$, which means that $P$ can  execute a transition and become $P'$.
There are three kinds of transitions:
\begin{itemize}
\item action transition: $\atransitionse\alpha$ where $\alpha$ is $g\envoi v$ or $g\recep x$;
\item silent transition: $\tautransitionse$, for internal transition;
\item delta transition: $\dtransitionse$, for successful termination.
\end{itemize}

For instance, the rule:
$$\frac{P\atransition{\alpha} P'}{Q\atransition{\beta}Q'}\hspace{10pt} K$$
can be read: if process $P$ can perform action $\alpha$ then become process $P'$, and if condition $K$ holds, then process $Q$ can perform action $\beta$ and become $Q'$.

In the following, $P, Q, P', Q', P_i$ and $P_i'$ are processes, $\alpha$ and $\alpha_i$ are actions,
$g$ is a communication gate, $v$ is a value, $x$ is a variable, and $c_j$ is a condition.

{\bf Successful termination}
$$
\frac{}{\term \dtransition \stopproc}
$$
{\bf Action Prefix}
$$
\frac{}{g \envoi v \prefix P \atransition{g\envoi v} P}\hspace{10pt}v\in\n
$$
$$
\frac{}{g \recep x \prefix P \atransition{g\recep x} P}
$$
{\bf Sequential composition}
$$
\frac{P \atransition\alpha P'}{P\seq Q \atransition\alpha P'\seq Q}
\hspace{10pt} \alpha \neq \delta
$$
$$
\frac{P \dtransition P'}{P\seq Q \tautransition Q}
$$
{\bf Parallel composition}
$$
\frac{P \atransition\alpha P'}{P\para Q \atransition\alpha P'\para Q}
\hspace{10pt} \alpha \neq \delta
$$
$$
\frac{Q \atransition\alpha Q'}{P\para Q \atransition\alpha P\para Q'}
\hspace{10pt} \alpha \neq \delta
$$
$$
\frac{P\atransition{g\envoi v}P' \hspace{10pt} Q\atransition{g\recep x}Q'}
{P\para Q \tautransition P'\para Q'[x\leftarrow v]}
$$
$$
\frac{P\atransition{g\recep x}P' \hspace{10pt} Q\atransition{g\envoi v}Q'}
{P\para Q \tautransition P'[x\leftarrow v]\para Q'}
$$
$$
\frac{P\dtransition P'\hspace{20pt} Q\dtransition Q'}
{P\para Q \dtransition \stopproc}
$$
{\bf Conditional choice}
$$
\frac{P_i \atransition{\alpha_i} P_i'}{\bigcond{j}{c_j}{P_j} \atransition{\alpha_i} P_i' }
\hspace{10 pt} c_i
$$
{\bf Restriction}
$$
\frac{P\atransition{\alpha}P'}{P\restrict L \atransition\alpha P'\restrict L}
\hspace{10pt}
\begin{array}{c}
\alpha=\tau \vee \alpha = \delta\\
\vee (\alpha = g[\envoi v \mbox{or}\recep x] \wedge g\not\in L)
\end{array}
$$

\section{The quantum process algebra}
\label{annexQPA}

\subsection{Syntax}

\begin{supertabular}{lcl}
\it elem\_cond & $\syntaxdef$ &\it variable $|$ value \\
\it comp & $\syntaxdef$ & $\pmb= | \pmb\neq | \pmb\le | \pmb\ge | \pmb< | \pmb>$ \\
\it cond & $\syntaxdef$ & \it elem\_cond comp elem\_cond\\
&&\\
\it var\_list & $\syntaxdef$ &\it variable $\{\pmb,$ variable $\}^*$ \\
&&\\
\it unitary\_operator & $\syntaxdef$ & $ CNot\ |\ H\ |\ I\ |\ X\ |\ Y\ |\ Z\ | \ldots$ \\
\it observable & $\syntaxdef$ & $M_{std}\ | \ldots$ \\
&&\\
\it communication & $\syntaxdef$ & \it gate {\bf !} value \\
                  &$|$& \it gate {\bf !} variable\\
                  &$|$&\it gate {\bf !} observable $\pmb [$ var\_list $\pmb ]$\\
 		&$|$& \it gate {\bf ?} variable  \\
\it unit\_transf         & $\syntaxdef$ & \it unitary\_operator $\pmb [$ var\_list $\pmb ]$ \\
\it measurement    & $\syntaxdef$ & \it observable $\pmb [$ var\_list $\pmb ]$\\
\it action & $\syntaxdef$ & \it communication \\
		&$|$&\it unit\_transf\\
		&$|$&\it measurement \\
&&\\
\it decl\_var & $\syntaxdef$ & \it $\pmb{\theta [}$ var\_list $\pmb ]\  |\ \pmb{\kappa [}$ var\_list $\pmb ]$\\
&&\\
\it process & $\syntaxdef$ & \it $\pmb\stopproc$\\
                  &$|$& \it $\pmb\term$ \\
                  &$|$& \it process\_name $[\pmb [$ var\_list $\pmb ]]$\\
                  &$|$& \it $\{$decl\_var$\}^+\pmb:$ process $\pmb\point$ \\
                  &$|$& \it action $\pmb\prefix$ process\\
                  &$|$& \it  process $\pmb\seq$ process\\
                  &$|$& \it  process $\pmb\para$ process\\
                  &$|$& \it  $\{\pmb\rect$ cond $\pmb\rightarrow$ process $\}^+$\\
                  &$|$& \it  process $\pmb | \pmb\{$ gate $\{$ ,gate $\}^*\pmb\}$ \\
&&\\
\it proc\_decl & $\syntaxdef$ & \it process\_name $\displaystyle\pmb{\procdef}$ process\\
\end{supertabular}

\subsection{Main inference rules of the semantics}

With respect to appendix \ref{annexSemCPA}, two new kinds of transitions have been added:
\begin{itemize}
\item declaration transition: $\transitionse$, for variable declaration;
\item probabilistic transition: $\ptransitionse p$, where $p$ is a probability.
\end{itemize}

In the following, $C$, $C'$ or $C_i$ represent contexts and $S_i$, a process state. 

\noindent
{\bf Variable declaration}
$$
\frac{}{\declcvar{x_1,\ldots,x_n}\ \declqvarp{y_1,\ldots,y_m}  P \point \spcontexte{C}
\transition P \point \spcontexte{C'}}
$$
with $C = \contexteqcq$, $C' = \contexte{s'}{q}{\ket\psi}{f}$ and\\
$s' = \ensemble{(x_1,\nattype),\ldots,(x_n,\nattype),
(y_1,\qubittype),\ldots,(y_m,\qubittype)}\pileajout s$

\noindent
{\bf Successful termination}
$$
\frac{}{\term\spcontexte C \dtransition \stopproc\spcontexte C}
\hspace{10pt}C\contextestable
$$
{\bf End of scope of variables}
$$
\frac{P \spcontexte{C} \divtransition P' \spcontexte{C'}}
{P \point \spcontexte{C} \divtransition P'\point \spcontexte{C'}}
$$
where $\divtransitionse$ stands for any of the transitions: $\atransitionse\alpha$ with $\alpha$ an action, $\tautransitionse$, or $\transitionse$.
$$
\frac{P \spcontexte{C} \dtransition P' \scontexte{e \pileajout s}{q}{\ket{\psi}}{f}}
{P \point \spcontexte{C} \dtransition \stopproc\ 
\scontexte{s}{q[e \leftarrow *]}{\ket\psi}{\domrestrict{f}{\varpile s}}}
$$
{\bf Action Prefix}
$$
\frac{}{g \envoi v \prefix P \spcontexte C \atransition{g\envoi v} P \spcontexte C}
\hspace{10pt} v \in \n,\ C\contextestable
$$
$$
\frac{}{g\envoi{x} \prefix P \spcontexte C \atransition{g\envoi{x}} P \spcontexte C} 
$$
where $C=\contexteqcq$, $x \in \varpile s$ and $x \in \dom f$
$$
\frac{}{g \recep x\prefix P \spcontexte C \atransition{g\recep x} P\spcontexte C}
$$
where $C=\contexteqcq$, $x \in \varpile s$
$$
\frac{}{U[x_1,\ldots,x_n]\prefix P \spcontexte C \tautransition
P \spcontexte{C'}}
$$
where
\begin{itemize}
\item $C=\contexteqcq$, $C'=\contexte{s}{q}{\ket{\psi'}}{f}$
\item  $U\in \transfu$, $x_1,\ldots,x_n \in \varpile s$, and $x_1,\ldots,x_n \in q$ 
\item $\ket{\psi'} = \Pi^t.(U\otimes I^{\otimes k}).\Pi\ket\psi$
\item $\Pi$ is the permutation matrix which places the $x_i$'s at the head of $q$
and $\Pi^t$ is the transpose of $\Pi$
\item $k = \tailleseq q - n\ $
\item $I^{\otimes k} = \underbrace{I \otimes \cdots \otimes I}_k$
\end{itemize}
$$
\frac{}{M[x_1,\ldots,x_n]\prefix P \spcontexte C \tautransition
P\scontexteprob{p_i}{s}{q}{\ket{\psi_i}}{f}}
$$
with
\begin{itemize}
\item $C=\contexteqcq$ (which implies $C\contextestable$)
\item $x_1,\ldots,x_n \in \varpile s$ and $x_1,\ldots,x_n \in q$ 
\item  $M\in\obs$  with $\sum_{i} \lambda_i P_i$ as spectral decomposition
\item $p_i = \langle \psi | \Pi^t (P_i \otimes I^{\otimes k}) \Pi | \psi \rangle$
\item $\displaystyle \ket{\psi_i} =  \frac{\Pi^t (P_i \otimes I^{\otimes k}) \Pi \ket{\psi}}{\sqrt{p_i}}$
\item $\Pi$ is the permutation matrix which places the $x_i$'s at the head of $q$
and $\Pi^t$ is the transpose of $\Pi$
\item $k = \tailleseq q - n\ $
\end{itemize}
$$
\frac{}{g \envoi {M[x_1,\ldots,x_n]} \prefix P \spcontexte C \tautransition
(g\envoi y\prefix \term \point)\seq P\spcontexteprob{p_i}{C_i}}
$$
where
\begin{itemize}
\item $y$ is a new variable
\item $C = \contexteqcq$ (which implies $C\contextestable$)
\item $C_i = \contexte{\ensemble{(y,\nattype)}\pileajout s}{q}{\ket{\psi_i}}
{\surcharge f \ensemble{y\mapsto\lambda_i}}$
\item and the conditions are the same as in the rule without communication.
\end{itemize}

\noindent
{\bf Sequential composition}
$$
\frac{P \spcontexte{C} \divtransition P' \spcontexte{C'}}
{P \seq Q \spcontexte{C} \divtransition P'\seq Q \spcontexte{C'}}
$$
where $\divtransitionse$ stands for any of the transitions~: $\atransitionse\alpha$ with $\alpha$ an action different from $\delta$, $\tautransitionse$, or $\transitionse$.
$$
\frac{P \spcontexte{C} \dtransition P' \spcontexte{C'}}{P \seq Q\spcontexte{C} \tautransition Q \spcontexte{C'}}
$$
{\bf Parallel composition}

In the rules for parallel composition, $C$, $C_P$ and $C_Q$ are defined as:
\begin{itemize}
\item $C = \contexte{(s_P\para s_Q)\pileajout s}{q}{\ket{\psi}}{f}$
\item $C_P = \contexte{s_P\pileconcat s}{q}{\ket{\psi}}{f}$
\item $C_Q = \contexte{s_Q\pileconcat s}{q}{\ket{\psi}}{f}$
\end{itemize}

In the definition of $C$, the operator $\para$ permits to build a cactus stack (see paragraph \ref{subsecQuantVar}).
In the cactus stack $(s_P\para s_Q)\pileajout s$ of the process $P\para Q$, the names in $s$
correspond to variables shared by $P$ and $Q$ whereas the names in $s_P$ (resp. $s_Q$) correspond to variables declared in $P$ (resp. $Q$).
$$
\frac{P \spcontexte{C_P} \divtransition P' \spcontexte{C_P'}}
            {P \para Q \spcontexte{C} \divtransition
              P' \para Q \spcontexte{C'}}
$$
where
\begin{itemize}
\item $\divtransitionse$ stands for one of those transitions~: $\atransitionse\alpha$ with $\alpha$ an action and $\alpha \neq \delta$, $\tautransitionse$, $\transitionse$

\item If $C_P' =\contexte{s'}{q'}{\ket{\psi'}}{f'}$ then
$C' = \contexte{(s_P'\para s_Q)\pileajout s}{q'}{\ket{\psi'}}{f'}$\\
with $s_P'$ such that $s' = s_P'\pileconcat s$ ($P$ can neither add to nor remove variables from $s$)

\item If $C_P' =\contexteprob{p_i}{s_i'}{q_i'}{\ket{\psi_i'}}{f_i'}$\\
then $C' = \contexteprob{p_i}{({s_P}_i'\para s_Q)\pileajout s}{q_i'}{\ket{\psi_i'}}{f_i'}$
with ${s_P}_i'$ such that $s_i' = {s_P}_i'\pileconcat s$
\end{itemize}
$$
\frac{P \spcontexte{C_P} \atransition{g\envoi v} P' \spcontexte{C_P'}
             \hspace{15pt}
              Q \spcontexte{C_Q} \atransition{g\recep x} Q' \spcontexte{C_Q'}}
            {P \para Q \spcontexte{C} \tautransition P' \para Q' \spcontexte{C'}}
$$
where
\begin{itemize}
\item $x \in \varpile s \cup \varpile{s_Q}$ and $v \in \n$
\item  If $x$ is of type $\nattype$, then:\\
$C' = \contexte{(s_P\para s_Q)\pileajout s} {q}{\ket{\psi}}{\surcharge f \ensemble{x\mapsto v}}$
\item  If $x$ is of type $\qubittype$, then:
$x\not\in q$, $v\in\ensemble{0,1}$\\
and $C' = \contexte{(s_P\para s_Q)\pileajout s} {x.q}{\ket v\otimes\ket{\psi}}{f}$
\end{itemize}
$$
\frac{P \spcontexte{C_P} \dtransition P' \spcontexte{C_P'}
             \hspace{20pt}
              Q \spcontexte{C_Q} \dtransition Q' \spcontexte{C_Q'}}
            {P \para Q \spcontexte{C} \dtransition \stopproc \spcontexte{C'}}
$$
with
$C' = \contexte{s}{q[(\varpile{s_P}\cup\varpile{s_Q})\leftarrow *]}
				{\ket{\psi}}	{\domrestrict{f}{\varpile{s}}}$

\noindent
{\bf Probabilistic contexts}
$$
\frac{}{P\spcontexteprob{p_i}{C_i} \ptransition{p_i} P\spcontexte{C_i}}
\mbox{ where } \sum_j p_j = 1
$$


\end{document}